\documentclass[english,a4paper,twocolumn,10pt]{article}
\title{\textbf{Account of the self-interaction energy  correction in the first principles calculation of the fundamental absorption edge spectrum of LiCl}}
\usepackage[T1]{fontenc}
\usepackage{lipsum}
\usepackage{geometry}
\usepackage{multicol}
\usepackage{graphicx}
\usepackage{amssymb}
\usepackage{babel}
\usepackage{amsmath}
\usepackage{authblk}
\author{M.A.Bunin\thanks{bunin.m.a@gmail.com},  and I.I Geguzin\\
Research Institute of Physics, Southern Federal University, Rostov-on-Don, Russia}
\begin{document}
\maketitle
\nobreak
	Within the framework of the local electron density functional theory, an ab-initio method is proposed that takes into account the self-interaction energy correction (SIC) for the crystal potential. The principle of dividing the unit cell into regions remained the same as for the ground-state potential. The expression for the self-consistent muffin-tin-SIC potential satisfies the condition of cell electroneutrality. This scheme was applied to calculate the spectrum of the fundamental absorption edge of LiCl. The obtained interband transition energies were not required for fitting to experimental values. In the calculated spectrum, the shape and position of the peaks above ~2.5 eV from the edge agree fairly well with the measured ones, which allows their intensity in this region to be attributed primarily to interband transitions. Closer to the edge, electron-hole interaction significantly alters the ground state, which must be taken into account when calculating the spectrum shape. 
	\\
		
\section{Introduction}
Calculations of the spectrum of elementary excitations of wide-bandgap dielectrics, such as alkali halide crystals (AHC), near the fundamental absorption edge using density functional theory (DFT) methods have not yet yielded satisfactory results.\footnote{This work, based on the results obtained in (M.A.Bunin, I.I Geguzin, Raschet opticheskogo pogloshcheniya LiCl iz pervykh printsipov s uchetom energii samodeystviya. Fizika Tverdogo Tela. 36, (1994) 1900-1909 (\textit{Calculation of the optical absorption of LiCl from first principles with account of the self interaction energy})), is inspired by the renewed interest in studying the role of self-interaction energy in understanding the practically important properties of wide-bandgap oxides. Given the some issues  are still considered, this text concerns the state of the problem at the time of obtaining the main results. This allows a clearer assessment of the existing achievements and the possibilities for further development.}. It was believed that the main problems is the incorrect description of the exchange-correlation contributions to the energy of the excited optical state, which did not take into account the nonlocality of the crystal potential. DFT can produce quite acceptable results as long as these contributions are small. Thus, for the fundamental absorption edge in a LiH crystal containing only 4 electrons in a unit cell, the calculation of optical spectra by the DFT method with Slater exchange reproduces well the experimental curves, the values of $\epsilon_{\infty}$ and of the band gap width $E_{g}$ [1]. The optical spectra of this AHC are  determined mainly by interband transitions. The electron-hole interaction transforms only a narrow ($\sim$0.3 eV) region near the edge and does not affect the rest part of the spectrum [1].

The example of lithium hydride shows that the one-electron \textit{ab-initio} approach can explain the optical spectra of AHCs. In contrast, the one-electron calculations from first principles known from the literature (see references in [2] and [3]) failed to do so. The reason is well known: the incorrect description of the nonlocal nature of the exchange-correlation coupling in the expression for the total energy of the local electron density appriximation (LDA) ground state (GS) gives an erroneous description of the excited optical state, which leads to an underestimated band gap of the AHC. So, the maim reason is the obviously improper representation of the crystal potential for excited optical state.  However, as the potential used becomes closer to an adequate representation of the considered physical process, the result of calculations can approach the experimental data. Then the first problem we have to solve is to find a suitable scheme that describes the optical excited state in LDA.

The GS density functional describes the probe electron interaction with all N electrons of the unit cell. In contrast to this, in the final state of the optical transition, the probe electron occupies conduction band states, interacting with N-1 electrons and a hole in the valence band. The hole can be accounted, for example, by methods of the defects theory. Thus, this description of the ground state of the optical transition formally resembles one of the methods that takes into account the self-interaction correction (SIC) to the ground state energy in LDA. This is mentioned in the review [4] and articles [1, $5-8$].

The role of the self-interaction correction energy in the theory of physical properties of materials studied in [9, 10] revealed that the complexity of its calculating in wide-bandgap dielectrics made it possible to study only the details of the main trends of the elementary excitations relative to the GS- spectrum [5, 9]. Accounting the self-interaction in the basis of localized orbitals of the LCAO method, the densities of states of six alkali halide crystals were calculated [11]. The SIC-accounted potential in the LMTO method when calculating the valence state densities of MgO and LiF in [12] demonstrated the improved gaps compared to the LDA-GS values. Unfortunately, the known examples consider only the gap value which corresponds to only one point in the entire spectrum.

In this paper, the idea proposed by Liberman [13] is used to construct the potential for LDA band calculations taking into account the SIC contribution. In first-principles one-electron methods of band calculations, such as KKR and APW, this has not been used before. The reason for this is the lack of such convenient methods for calculating the potential as in [14,15], which previously ensured progress in the electronic structure calculations of crystals. 
  
The aim of this work is to study the possibility of one-electron description of optical spectra of wide-bandgap dielectrics using basisless calculation with a potential that accounts the self-interaction energy contribution. An expression for the self-consistent muffin-tin-SIC potential is obtained. It satisfies the requirement of electroneutrality of the unit cell, which is important for the correct calculation of energies. For convenience of calculations and comparison of results obtained in the SIC- and GS- approaches, the method of dividing the cell space remained the same for both of them.

\section{MT-SIC potential}
Let's obtain expressions for the MT-SIC potential in a crystal which unit cell contains N electrons and M atoms. For its ground state, the total energy functional can be written as:

	\begin{multline}
		E^{0}[\rho(\vec r)]=T[\rho(\vec r)]+\iint\limits_{\Omega}\frac{\rho(\vec r ')\rho(\vec r)}{|\vec r '-\vec r|}d\vec r 'd\vec r+ \\
		\int\limits_{\Omega}\rho(\vec r)\varepsilon_{\mathit{xc}}[\rho(\vec r)]d\vec r-N{\sum _{\mathit{n,s,s'}}}^{'}\frac{Z_s Z_{s'}}{|\vec r_n '-\vec r_{s'}-\vec r_s|}
		\end{multline}
where the first term is the kinetic energy; the second is the Coulomb energy of electrons (this includes the contribution of the electron-nuclear interaction); the third is the exchange-correlation energy, which is often calculated in a local density approximation; the fourth is the Madelung contribution from the field of point charges of ions located at the lattice sites (the prime means that this sum already takes into account the contribution of the self-interaction, i.e. $s \ne s'$). Varying this functional under conditions of the cell electroneutrality:	
	
	\begin{equation}
		\int\limits_{\Omega}\rho\vec(r)d\vec r=N
	\end{equation}
	provides well-known expressions for the \textit{mt} potential of the  crystal's ground state [11,12]. 
\\

In constructing the potential that accounts the SIC, we combine the \textit{mt}-model with the generalized idea of including SIC for atoms and monatomic crystals proposed by Lieberman in 1970 [13]. That satisfies to the model proposed for the optical transition in the Introduction. In addition, the condition of electroneutrality of the unit cell must be satisfied. The partition of the unit cell space in the \textit{mt}-form  is convenient for relating to expressions obtained for the GS potential [11] and facilitates comparison with the results of previous calculations. In addition, this allows the use of existing calculation algorithms in the most convenient form. As notations, we will use the superscript \textit{mt} or \textit{sic} to distinguish the corresponding parameters.
\\

According to the \textit{mt}-model, the space of the unit cell $\Omega$ is divided into regions of two types: inside the contacting spheres (which radii is $\textit{R}^{mt}_{s}$) around each atomic site $\vec {R}_{s}$ and outside them ($\Omega^{mt}_{II}$). For further consideration, it is more convenient to assume some definite type of electron density distribution. Let's consider an ionic - type crystal (but the same can be done for others). Its main part of the electron density is localized near the atomic sites, being small in the interatomic region, so that in the interspherical region ($\Omega^{mt}_{II}$ in Fig. 1) the following condition must be satisfied:
	
	\begin{equation}
	\int\limits_{\Omega^{mt}_{II}}\rho(\vec r)d\vec r<1
\end{equation}
i.e. there is less than one electron outside the \textit{mt}-spheres. Continuing in this way, we divide the unit cell into parts such that one part contains one electron, and the other part contains the remaining N-1 electrons, which corresponds to the optical transition model proposed above. This can be done by inscribing into each \textit{mt}-sphere another one with the same center and smaller radii $\textit{R}^{sic}_{s}$ < $\textit{R}^{mt}_{s}$ . Ultimately, the cell space is divided into regions (Fig. 1): $\Omega^{sic}_{I}$ – the inner part of the spheres with radii $\textit{R}^{sic}_{s}$ (\textit{sic}-region), the outer part $\Omega^{sic}_{II}$, consisting of all spherical layers between the spheres $\textit{R}^{mt}_{s}$ and $\textit{R}^{sic}_{s}$ (\textit{s} = 1, ..., M), and the interspherical part $\Omega^{mt}_{I}$. In this case, within each \textit{mt}-region, the potential remains spherically symmetric. The total potential of the \textit{s}-atom of the cell can be represented as: 
	
	\begin{equation}
		v(\vec r)=\left\{
		\begin{array}{l}
		v_{1}^{s}(|\vec r-\vec {R_{s}}|),               |\vec r-\vec {R_{s}}|\le R_{s}^{sic} \\	
		v_{2}^{s}(|\vec r-\vec {R_{s}}|),   R_{s}^{sic}<|\vec r-\vec {R_{s}}|\le R_{s}^{mt}\\
		v_{3} = const,                       \vec r \in \Omega^{mt}_{II}
			\end{array}
	\right.
	\end{equation}
where $\vec{R}_{s}$ are the positions of the atoms in the cell. The values of $\textit{R}^{sic}_{s}$ are found from the condition:
	\begin{equation}
	\int\limits_{\Omega^{sic}_{II}}\rho(\vec r)d\vec r=1
\end{equation}
where the integration is perform over a region $\Omega^{sic}_{II}$, external to the inscribed spheres $\textit{R}^{sic}_{s}$.
	
\begin{figure}[!ht]
	\begin{center}
		\includegraphics[width=70 mm, keepaspectratio]
		{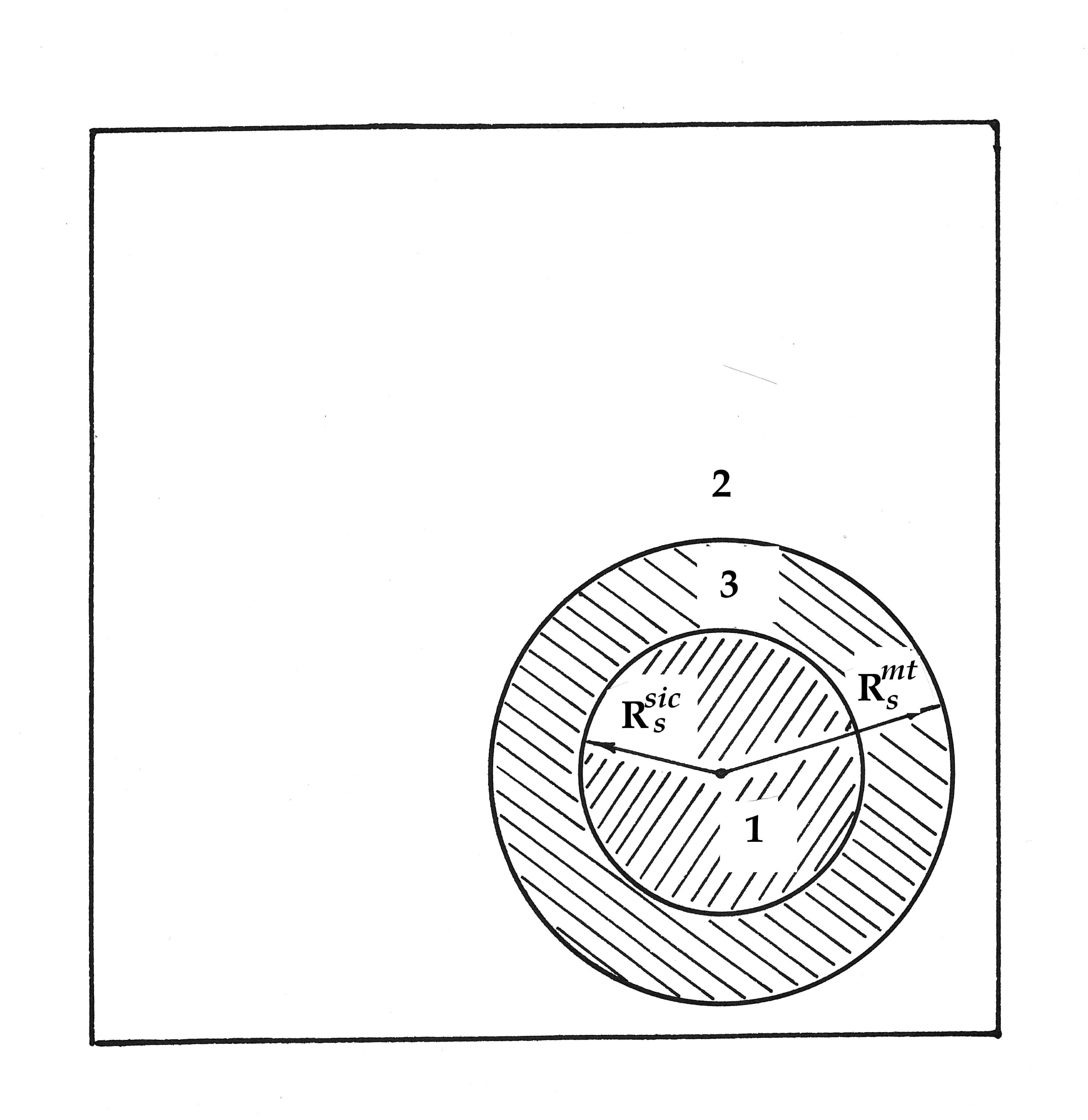}\caption{Partitioning of the unit cell space $\Omega$:  $(1)=\Omega_{\textit{s}}^{\textit{sic}}$; $ (2)=\Omega_{II}^{\textit{mt}}$;  $(1)+(3)=\Omega_{\textit{s}}^{\textit{mt}}$;  $\Omega_{I}^{\textit{sic}}=\sum_{s}\Omega_{s}^{\textit{sic}}$, $\Omega_{I}^{\textit{mt}}=\sum_{s}\Omega_{s}^{\textit{mt}}$ 
		}\label{Рис2}
	\end{center}
\end{figure}

The correction for self-interaction in functional (1) can be taken into account by subtracting the energies of the Coulomb and exchange interactions of the electron with itself from $E^{0}[\rho(\vec r)]$. To calculate these contributions, it is convenient to use the above-proposed method of dividing the cell space together with Lieberman's idea of constructing a potential that takes SIC into account. According to (5), there is a unit electronic charge in the $\Omega^{sic}_{II}$ region. The corresponding electron density is the sum of the contributions of the peripheral parts of the wave functions of all atoms in the cell and cannot be attributed to any specific electron shell. Then, in the contributions of the self-energy subtracted from $E^{0}[\rho(\vec r)]$, it is sufficient to restrict the integration over the $\Omega^{sic}_{II}$ region to obtain a functional with a correction for the self-interaction energy:
		\begin{multline}
		E^{sic}[\rho(\vec r)]=E^{0}[\rho(\vec r)]-\\
		\left\lbrace \iint\limits_{\Omega_{II}^{sic}}\frac{\rho(\vec r ')\rho(\vec r)}{|\vec r '-\vec r|}d\vec r 'd\vec r+	
		\int\limits_{\Omega_{II}^{sic}}\rho(\vec r)\varepsilon_{\mathit{xc}}[\rho(\vec r)]d\vec r \right\rbrace
		\end{multline}
	  
 This form of the self-interaction correction (6) is close to common expressions in [4,5] and is similar to those ones in [10]. Varying (6) under conditions (2) and (5) and with  condition of the unit cell electroneutrality  gives expressions for local representations $\textit{v}_{1}^{s}(r)$ , $\textit{v}_{1}^{s}(r)$  and $\textit{v}_{3}$ of the crystal potential accounted the self-interaction energy:
 
	 \begin{multline}
	v_{1}^{s}(r)=-\frac{2Z_{s}}{r}+v^{coul}_{s}(r)+v^{xc}_{s}(r)-\\2\sum_{s^{'}}{M_{ss^{'}}\Delta Z_{s^{'}}}+\lambda (R^{sic}_{s});\\
	v_{2}^{s}(r)=-\frac{2\Delta Z_{s}}{r}-2\sum_{s^{'}}{M_{s{s^{'}}}\Delta Z_{s^{'}}};\\  
	v_{3} = \left({\Omega^{mt}_{s}}\right)^{-1}\sum_{s}\left\lbrace 3\left( \frac{4\pi}{3}\right)^{1/3}\right.\times\\
	\left\lbrack{\Omega^{2/3}-\left({\Omega_{s}^{mt}}\right)^{2/3}}\right\rbrack\Delta Q_{s}-2\left.\sum_{s^{'}}M_{s s^{'}}\Delta Q_{s^{'}}\right\rbrace . 
	  \end{multline}
where $r\in \Omega^{mt}_{s}$; $\textit{s}$ is the number of an atom in a unit cell; $Z_{s}$ are the nuclei charges; $M_{ss^{'}}$ is a Madelung matrix; $\Delta Z_{s}$=$Z_{s}$ - $Q_{s}$ is the excess of positive charge in the corresponding sphere (\textit{sic} or \textit{mt});  $\Delta Q_{s}$ is the electron charge inside the spherical layer; the numerical value of $\lambda (R^{sic}_{s})$ is determined from the condition of potentials equality at the radius of the sphere $R^{sic}_{s}$; $v^{coul}_{s}(r)$  is the Coulomb potential; $v^{xc}_{s}(r)$  is the exchange-correlation contribution to the potential of the sphere \textit{s} (most often it is calculated in the statistical exchange approximation, which we will also use); $\Omega$ is the unit cell volume
	
	\section{The influence of self-interaction on the electronic structure of LiCl} 
		
The proposed method for constructing the MT-SIC potential is used to calculate the fundamental absorption edge spectra of LiCl. The ground state of LiCl is calculated by the APW method with the Slater's exchange (with the $ \alpha = 1$). The radii of the \textit{mt}-spheres are chosen based on the condition of a minimum electron density in the intersphere region (Table 1). In the SIC calculation, the GS potential was the starting one. Twelve iterations were required to achieve self-consistency with an electron density difference of 0.001 in two successive iterations. Since the valence electron density is concentrated mainly on chlorine ($\sim$ 98 $\% $ of the charge), it is advisable to include into the $\Omega^{sic}_{II}$ - region only the vicinity of this atom and $\Omega^{mt}_{II}$, and for Li make the radii of the \textit{mt}- and \textit{sic}-spheres the same. The calculation parameters for LiCl  and the electron charges of the \textit{mt}-spheres are given in Table 1.

\begin{table}[htbp]
	\centering
	\caption{Parameters of LiCl calculation and the electron charges in \textit{mt}-spheres.\\
}
\begin{tabular}{c|c|c|c|c}
	& \multicolumn{2}{c|}{Radii, a.u.} & \multicolumn{2}{c}{Charges}\\
	Approach  & Li & Cl & Li & Cl \\
	\hline
	GS & 1.757 & 3.100 & 2.04 & 17.57 \\
	SIC & $\large{\frac{1.757}{1.757}}$ & $\large{\frac{3.100}{2.868}}$ & 2.04 & 17.58 \\

\end{tabular}\\
\end{table}
		
The forbidden gap width is one of the important characteristics of the crystal. For the GS- potential, the calculation gives 7.47 eV. Taking SIC into account, we obtained 8.33 eV, which is better, but still less than the experimental value of 9.4 eV [16].

Since the calculation of optical spectra is based on the  momentum matrix elements $p_{if,\textbf{k}}$, it is advisable to compare their values calculated at different approaches: the GS and the SIC at different points of the Brillouin zone for different transition energies (it is convenient to compare their square values, Table 2). It is evident that the self-interaction correction significantly affects the $p_{if,\textbf{k}}$ values. For all energies and points of the Brillouin zone $(p_{if,\textbf{k}}^{SIC})^{2} >(p_{if,\textbf{k}}^{GS})^{2}$ (and not just for those presented in Table 2), the most significant changes occur near  the conduction band bottom (in optical spectra, this is the fundamental absorption edge $\omega_{0}$). As energy increases, the differences decrease but remain dependent on the $\textbf{k}$ -point.Thus, taking into account SIC leads to complex (in any case, nonlinear) changes in the $p_{if,\textbf{k}}$ values. Since the latter contain information both on the wave functions and on the local partial densities of states $N_{\tau}(E)$ (index $\tau$   denotes the local symmetry with respect to a particular atom of the cell), for greater clarity these factors should be separated. First, we considered the wave functions of states with angular momenta $l = 0, 1, 2$  at energies in the range of $\sim$ 2 Ry from the conduction band bottom, as well as at the top of the valence band. For both approaches, the amplitudes of the radial parts of the wave functions differ on average by (0.1-0.5)$\% $ and  in any case by no more than $1\% $. A comparison of the densities of states (their detailed analysis can be done in a separate work) shows that this factor is more significant. It manifests itself in an uneven shift of separate parts of $N_{\tau}(E)$ and a redistribution of their contributions. The nature of the changes depends on the symmetry of the states relative to the corresponding center (Li or Cl in our case).

\begin{table}[htbp]
\centering
\caption{
	Square values of the momentum matrix elements  
	$(p_{if,\textbf{k}}^{sic})^{2} = \left(-i\int\limits_{\Omega} {\psi{^{*}_{f\textbf {k}}(\vec r )}\nabla\psi{^{*}_{i\textbf{k}}(\vec r)}}d\vec r \right)^{2}$, Ry for\\ interband transitions $(\textbf{\textit{i}}\rightarrow\textbf{\textit{f}})$ in different points of the Brillouine zone.\\}	
\begin{tabular}{c|c|c|c}

	Point $\vec k$ $4\left(2\pi/a\right)$ & $\textbf{\textit{i}}\rightarrow\textbf{\textit{f}}$ &  $p^{SIC}_{if}{/}p^{GS}_{if}$ & $\omega^{SIC}_{if}{/}\omega^{GS}_{if}$\\
	\hline 	
	$\Gamma$ & 1 $\rightarrow$   4 & 0.182/0.169 & 0.612/0.549 \\
	& 1 $\rightarrow$   5 & 0.258/0.281 & 0.962/0.970 \\
	& 1 $\rightarrow$   9 & 0.236/0.231 & 1.146/1.165 \\ 
	X & 2 $\rightarrow$  4 & 0.303/0.295 & 0.780/0.766 \\
	& 1 $\rightarrow$  5 & 0.153/0.145 & 0.920/0.884 \\
	& 2 $\rightarrow$   5 & 0.176/0.166 & 0.81/0.779 \\
	& 2 $\rightarrow$   9 & 0.178/0.174 & 1.352/1.370 \\ 
	(321) & 3 $\rightarrow$   4 & 0.298/0.271 & 0.771/0.731 \\
	& 3 $\rightarrow$  8 & 0.130/0.116 & 1.295/1.257 \\
	& 1 $\rightarrow$  8 & 0.102/0.098 & 1.370/1.368 \\

\end{tabular}
\end{table}

For illustration, the $\textit{a}_{1g}$- and $\textit{e}_{g}$-  states densities of chlorine are shown in Fig. 2. From these curves, it can be concluded that in the SIC calculation the electronic structure changes, although the main features of the $\textit{N}_{\tau}(\textit{E})$ distribution of the ground state can be preserved. This result is completely consistent with the known property of self-interaction correction, which should have a greater effect on the energies of the states and a smaller effect on their wave functions.

\begin{figure}[!ht]
	\begin{center}
		\includegraphics[width=70 mm, keepaspectratio]
		{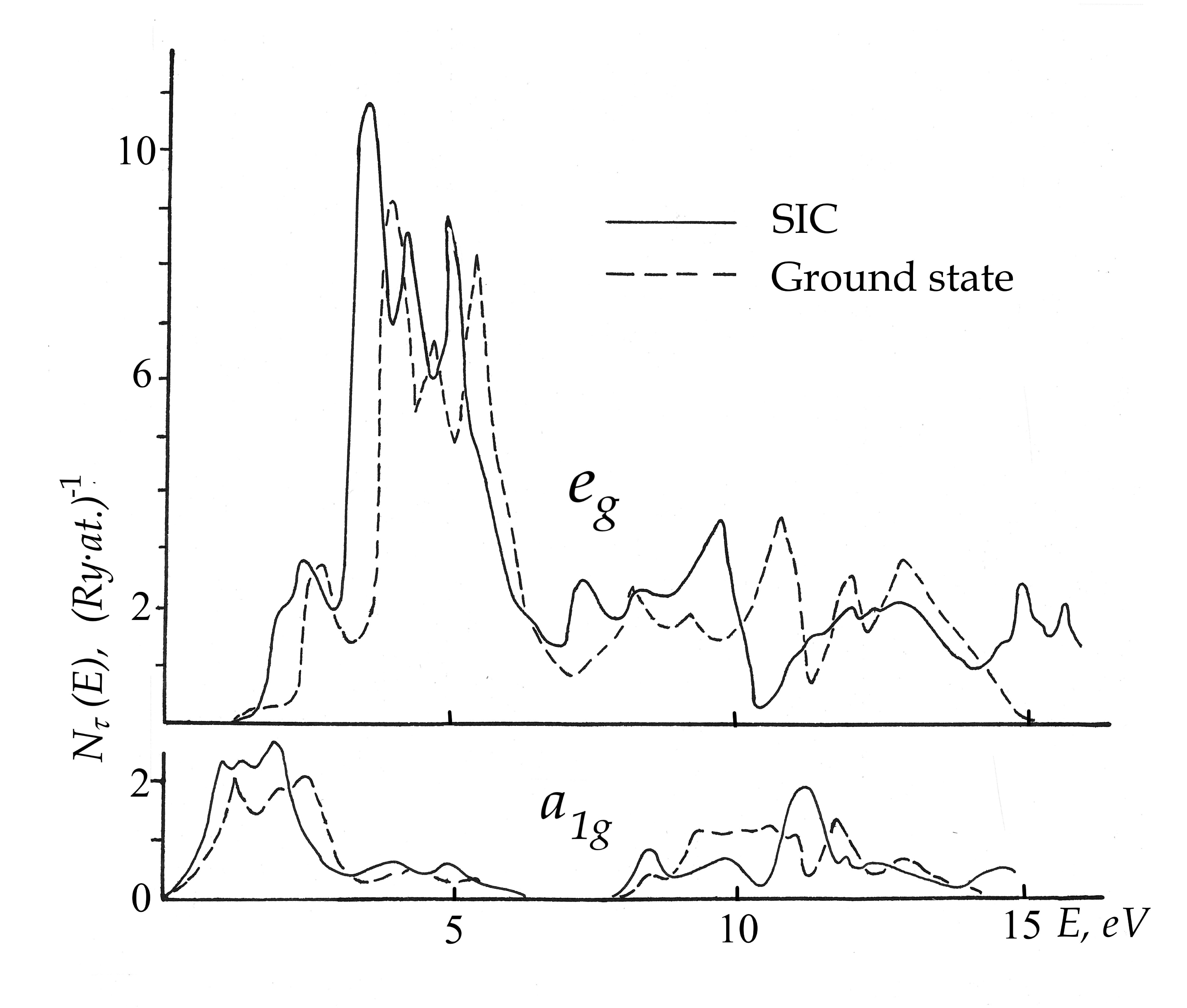}\caption{ Local densities of states for chlorine atom in LiCl, calculated for  $\textit{a}_{1g}$ (lower part) and $\textit{e}_{g}$ (upper part) states for SIC (solid line) and GS approaches (dashed line). Zero on the energy scale corresponds to the conduction band bottom.  
		}\label{Fig2}
	\end{center}
\end{figure}
		
	\section{Fundamental absorption edge of LiCl}	
		
	Of the experimental data on the optical functions of LiCl in a wide energy range, we know only the absorption spectrum $\mu(\omega)$ from the monograph [3]. In Fig. 3a, the absorption spectrum obtained from the SIC calculation is compared with the experimental one. In Fig. 3b the theoretical $\mu(\omega)$ are compared for SIC and GS approaches. No corrections were made to bring the $\omega$ values into agreement with the experiment, and lifetime smearing was not taken into account.
	
	\begin{figure}
		\begin{center}
			\includegraphics[width=80 mm, keepaspectratio]
			{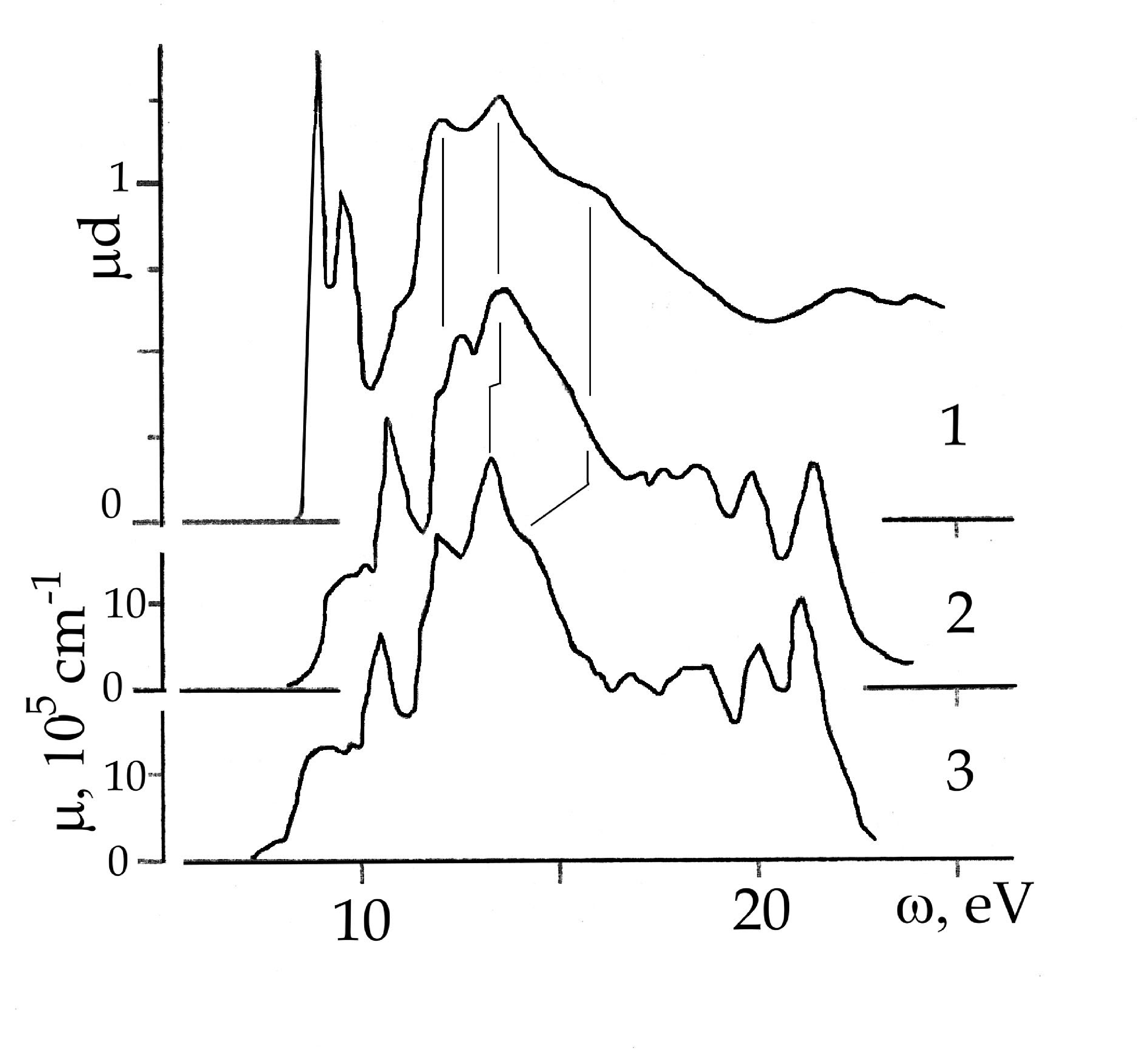}\caption{(\textbf{a})- The fundamental absorption edge spectrum $\mu(\omega)$ for LiCl: 1$-$ experimental spectrum from [3], 2$-$ SIC calculation, 3 $-$ GS calculation.  
			}\label{Fig3}
		\end{center}
	\end{figure}
	
	First, let us compare the calculated $\mu(\omega)$ for SIC and GS (Fig. 3(b)). The GS spectrum is shifted towards lower energies relative to the SIC spectrum. In addition, the intensities of the main features are distributed in it somewhat differently, although the similarity with the experimental curve in Fig. 3a is preserved. It is important to note that shifting any of the curves by a constant value does not lead to their coincidence. The reason for this is the difference in the matrix elements $p_{if,\textbf{k}}$ discussed above, and ultimately the different, although partly similar, electronic structure obtained by the two cosidered approaches.
	
	Now we can compare the results of the SIC calculation with the experiment, starting from the region away from the fundamental edge $\omega_{0}$.
	
	In the region $\omega > \omega_{0} + 2.5$ eV the calculated curve agrees well with the experimental one not only in the relative intensity of the spectral features but also (which is very important for assessing the quality of the results) in their position. The maximum in the region of $11-12$ eV corresponds to a peak in the experimental curve. In the region of $12-15$ eV the main maxima of both spectra practically coincide. At energies $\omega $ > 23 eV the drop in the calculated curve is due to a purely technical limitation of the upper boundary of the calculated spectrum. Overall, the agreement with the experiment is good for this frequency region. Then the LiCl optical absorption  for $\omega > \omega_{0} + 2.5$ eV is explained by interband transitions. It is important to note that the shift of the ground state spectrum $\mu(\omega)$ by a constant value, often used to replace the correction for self-interaction, improves agreement with experiment only in one narrow frequency range. Our calculation showed the need to take into account self-interaction to obtain transition energies and optical spectral shapes close to experiment in a wide frequency range, quite far from the edge.
	
	Let us consider the region near the threshold. In the interval $\omega_{0} < \omega <\omega_{0}+ 2.5$ eV, the electron-hole interaction strongly transforms the spectrum: instead of the stair in the calculated $\mu(\omega)$ the experimental curve shows two bright maxima, the first of which is most likely of exciton nature. The position of the second corresponds to a stair in the SIC spectrum $\mu(\omega)$. Its origin may be explained by calculating the transformation of the corresponding part of the density of states in the field of the hole.
	
	Thus, two regions can be distinguished in the considered spectrum. The first is near the edge, where the influence of the electron-hole interaction is significant, creating excitonic lines and changing the shape of the interband optical spectra. In LiCl, such an interval width is 2-2.5 eV compared to 0.2-0.3 eV in the optical spectrum of LiH [1]. This allows us to assume that the range of the $\omega $ values, where it is necessary to take into account the hole, depends on the number of electrons in the cell. It is possible that in AGCs with a higher atomic weight of the constituent components, the width of this region will be even greater. The hole can be taken into account using defect theory methods. In the second region, further from the edge, the shape of the optical spectrum is determined by interband transitions. To achieve agreement with experiment, self-interaction should be taken into account when constructing the crystal potential.	
\section{Conclusion}		 
	- A method is proposed that takes into account the correction for the self-interaction energy in the local electron density approximation. Minimization of this functional taking into account the requirement of electroneutrality of the cell yields an expression for the potential energy similar to the known ones for the ground state [14,15] and convenient for calculating the electronic structure from first principles using basisless \textit{mt}- methods such as KKR or APW.		
				
	-The calculation of LiCl showed that the effect of the self-interaction correction is manifested mainly in changes of the electronic structure details. At the same time, the peculiar features of the distribution of partial densities of states calculated for the potential of the ground state of the crystal can be preserved. 	
	
	-The suitability of the proposed scheme for calculating the optical spectra of wide-bandgap dielectrics is demonstrated using the example of optical absorption of LiCl. Taking into account the self-interaction energy allows us to explain the experiment without introducing corrections into the theory, such as an energy scale shift. This approach to the problem of the optical spectra of alkali-halide crystals agrees better with the experiment than the model based on the ground-state potential.
	
	- In the fundamental absorption edge spectrum of LiCl, two regions can be distinguished: the closest to the edge, where the spectrum of interband transitions is hardly changed by the electron-hole interaction, and the region far from the edge, where the spectrum is determined by interband transitions.\\

The authors are grateful to V.A. Lobach for kindly providing copies of his works.\\

The work was partially carried out with the financial support of the Ministry of Education and Science of the Russian Federation (GZ in the field of scientific activity 2023, Project No.FENW-2023-0015).

\end{document}